\documentclass{article}
\usepackage{spconf,amsmath,graphicx}
\usepackage{booktabs} 
\usepackage{xcolor}

\title{\vspace{-2mm}Stable Audio Open}
\name{Zach Evans\hspace{5mm}Julian D. Parker\hspace{5mm}CJ Carr\hspace{5mm}Zack Zukowski\hspace{5mm}Josiah Taylor\hspace{5mm}Jordi Pons}
\address{Stability AI}

\begin{document}
\ninept
\maketitle
\begin{abstract}
Open generative models are vitally important for the community, allowing for fine-tunes and serving as baselines when presenting new models.
However, most current text-to-audio models are private and not accessible for artists and researchers to build upon. Here we describe the architecture and training process of a new open-weights text-to-audio model trained with Creative Commons data. Our evaluation shows that the model's performance is competitive with the state-of-the-art across various metrics. Notably, the reported $\text{FD}_{openl3}$ results (measuring the realism of the generations) showcase its potential for high-quality stereo sound synthesis at 44.1kHz.

\end{abstract}
\begin{keywords}
Stable Audio Open, Latent Diffusion, Audio.
\end{keywords}
\section{Introduction}
\label{sec:intro}
A significant amount of ongoing research focuses on text-conditioned generative audio models~\cite{agostinelli2023musiclm, kreuk2022audiogen,liu2023audioldm, liu2023audioldm2, stableaudio, evans2024long}. Yet, many models lack public weights or are only available behind private APIs~\cite{agostinelli2023musiclm, stableaudio, evans2024long}, limiting their usefulness as the foundation for further research and artistic creation. 
Further, the licenses of the audio used for training public models are often not fully documented.
For example, AudioGen~\cite{kreuk2022audiogen} or AudioLDM~\cite{liu2023audioldm, liu2023audioldm2} are trained on a mix of public datasets that include AudioSet~\cite{gemmeke2017audio}, and we are unaware of any disclosed licenses for AudioSet's audio. MusicGen~\cite{copet2023simple}, on the other hand, is an open model with well-documented training data and licences, trained exclusively on licensed\footnote[1]{Documented in: https://github.com/facebookresearch/audiocraft/blob/\newline main/model\_cards/MUSICGEN\_MODEL\_CARD.md} copyrighted data. Finally, current open models are not competitive against the state-of-the-art in terms of quality, coherent generation over long sequences, and inference speed~\cite{stableaudio, evans2024long}. 
Given that, our goal is to release a text-conditioned generative model for non-speech audio based on the following:
\begin{itemize}
    \item Trained only on Creative Commons (CC) licensed audio. 
    \item Publicly available model weights and code, along with the attributions needed for the used data, to facilitate open research.
    \item State-of-the-art sound quality generation at 44.1kHz stereo.
\end{itemize}

\noindent 
While this data choice may limit our model's capabilities (specially for text-to-music as noted in section~\ref{sec:limitations}) it facilitates transparent data practices at the time of releasing the model. 
This manuscript describes how these goals were achieved, including the description of our architecture (section~\ref{sec:architecture}), how we handle training data and ensure that such audio is not copyrighted (section~\ref{sec:training_data}), and how our model was trained (section~\ref{sec:training}). 
In section~\ref{sec:evaluation} we evaluate the resulting model and its parts (the generative model and the autoencoder) by a number of standard metrics and study whether it exhibits memorization of the training data. We also show that our model can run on consumer-grade GPUs. Our aim is to further improve current best practices for open model releases, with an emphasis on evaluation, data transparency, and accessibility for artists and scholars.

\newpage

\section{architecture}\label{sec:architecture}

Our latent diffusion model generates variable-length (up to 47s) stereo audio at 44.1kHz from text prompts. 
It consists of 3 parts: an autoencoder (156M parameters) that compresses waveforms into a manageable sequence length, a T5-based text embedding~\cite{raffel2020exploring} for text conditioning (109M parameters), and a transformer-based diffusion model (DiT of 1057M parameters) that operates in the latent space of the autoencoder. Our model is a variant of Stable Audio 2.0~\cite{evans2024long} that is trained on CC data. Its architecture is similar except that it uses T5 \cite{raffel2020exploring} text conditioning instead of CLAP~\cite{wu2023large}. 
The model's exact parameterization and its weights are available online.\footnote[2]{https://huggingface.co/stabilityai/stable-audio-open-1.0/}


\subsection{Autoencoder}

The (variational) autoencoder operates on raw waveforms. The encoder processes such waveforms with 5 convolutional blocks, each performing downsampling and channel expansion via strided convolutions. Before each downsampling block, we employ a series of ResNet-like layers using dilated convolutions and Snake~\cite{snake} activation functions for further processing. The bottleneck of the autoencoder is parameterized as a variational autoencoder with a latent size of 64. The decoder is an inversion of the encoder structure, employing transposed strided convolutions for upsampling, with channel contraction before each upsampling block. All convolutions are weight-normalised and the output of the decoder does not include a $tanh()$ as we found it introduced harmonic distortion.

\subsection{Diffusion-transformer (DiT)}\label{sec:dit}
Our generative model is a diffusion-transformer~(DiT)~\cite{evans2024long,Peebles_2023_ICCV,apple} that follows a standard structure with stacked blocks consisting of serially connected attention layers and gated multi-layer perceptrons~(MLPs), with skip connections around each. We employ bias-less layer normalization at the input to both the attention layer and the MLP. The key and query inputs to the attention layer have rotary positional embeddings~\cite{rope} applied to only half of the embeddings. {Each transformer block also contains a cross-attention layer to incorporate conditioning.} Linear {mappings} are used at the input and output of the transformer to translate from the autoencoder latent dimension to the embedding dimension of the transformer. 
Efficient block-wise attention~\cite{dao2022flashattention} and gradient checkpointing~\cite{chen2016training} are employed to reduce the computational and memory requirements.

The DiT is conditioned by 3 signals: \textit{text} enabling language control, \textit{timing} enabling variable-length generation, and \textit{timestep} signaling the current diffusion timestep. Text conditioning is provided by the pretrained T5-base encoder \cite{raffel2020exploring}. Timestep conditioning~\cite{stableaudio,dhariwal2020jukebox} goes through sinusoidal embeddings~\cite{ho2020denoising}. {Conditioning is introduced via cross-attention or via prepending conditioning signals to the input.
Cross-attention includes {timing} and {text} conditioning. Prepend conditioning includes {timing} and  {timestep} conditioning.}

\subsection{Variable-length audio generation}\label{sec:variable}

As natural audio can be of various lengths, our model supports variable-length audio generation within a specified window (e.g., 47s) by relying on the timing condition to fill the signal up to the specified length. The model is trained to fill the rest with silence. To present variable-length audios (shorter than the window length) to the end-users, one can easily trim the appended silence. We adopt this strategy, as it has previously shown its effectiveness~\cite{stableaudio, evans2024long}.

\section{Training Data}
\label{sec:training_data}

Our dataset consists of CC recordings from Freesound and the Free Music Archive (FMA).
We conducted an analysis to ensure no copyrighted content was in our training data. To that end, we first identified music recordings in Freesound using the PANNs~\cite{kong2020panns} tagger. 
The identified music activated music-related tags for at least 30 sec (threshold of 0.15).
This threshold was set with FMA music examples and ensuring no false negatives were present. The identified music was sent to a trusted content detection company to ensure the absence of copyrighted music. The flagged copyrighted music was subsequently removed from our training set. Most of the removed content was field recordings in which copyrighted music was in the background. Following this procedure, we are left with 266,324 CC0, 194,840 CC-BY, and 11,454 CC Sampling+ audio recordings.

We also conducted an analysis to ensure no copyrighted content was present in FMA's subset. In this case, the procedure was different because the FMA subset consists of only music. We did a metadata search against a large database of copyrighted music and flagged any potential match to be reviewed by humans. After this process, we ended up with 8,967 CC-BY and 4,907 CC0 tracks.

This led to a dataset with 486,492 recordings {(7,300~h)}, where 472,618 {(6,330 h)} are from Freesound and 13,874 {(970~h)} from FMA, all licensed under CC-0, CC-BY, or CC-Sampling+. The most common tags in those Freesound recordings are \textit{single-note}, \textit{synthesizer}, \textit{field-recording}, \textit{drum}, \textit{loop}, \textit{ambient} and in FMA are \textit{instrumental}, \textit{electronic}, \textit{experimental}, \textit{soundtrack}, \textit{ambient}. This data is used to train most of our system (autoencoder and DiT, not the publicly available T5-base \cite{raffel2020exploring} that was pretrained) from scratch.

\subsection{Autoencoder training data}

We gathered 5 sec chunks of diverse, high fidelity audio. First, we gathered up to $\times$3 (random) chunks for each Freesound audio, to ensure diversity and avoid oversampling long recordings. Then, we gathered additional (random) chunks from a high fidelity subset, including all FMA tracks and a subset of stereo and full-band Freesound audio (55,314). Such high fidelity Freesound audio were recorded at 48kHz and verified to contain energy in high frequencies. FMA tracks were variable-bitrate MP3s encoded at 44.1kHz. {Note that high fidelity Freesound recordings were sampled twice.}

\begin{table*}[]
\centering
\begin{tabular}{lcccccc}
\toprule
                          &  & output  & & &  \\
                          & channels/sr &  length & $\text{FD}_{openl3}$ $\downarrow$ & $\text{KL}_{passt}$ $\downarrow$ & $\text{CLAP}_{score}$ $\uparrow$  \\ \midrule             
AudioLDM2-48kHz~\cite{liu2023audioldm2}  & 1/{48kHz} & 10 sec &  {101.11} &  2.04  &  0.37    \\ 
{AudioLDM2-large}~\cite{liu2023audioldm2}  & 1/16kHz & 10 sec &  170.31  &  1.57 &  0.41   \\  
{AudioGen-medium}~\cite{audiogen}   & 1/16kHz  & 10 sec &  186.53 & {{1.42}} &  {{0.45}}  \\ \midrule
{Stable Audio 1.0}~\cite{stableaudio}   & {2}/44.1kHz & 95 sec $^\dagger$ &  {103.66} & 2.89 &  0.24  \\
{Stable Audio 2.0}~\cite{evans2024long}   & {2}/44.1kHz & 190 sec $^\dagger$ &   116.14 & 2.67 & 0.24  \\
{Stable Audio 2.0}~\cite{evans2024long}   & {2}/44.1kHz & 285 sec $^\dagger$ &   110.62 & 2.70 & 0.23  \\
{Stable Audio Open}    & {2}/44.1kHz & 47 sec $^\dagger$ &  \textbf{78.24} & \textbf{2.14} & \textbf{0.29}  \\ 
\bottomrule
\end{tabular}
\caption{\textit{AudioCaps Dataset (with sounds, not music).} Stable Audio Open outperforms comparable baselines, showcasing its potential in synthesizing sounds and field recordings. Since AudioCaps is a subset of AudioSet, our results might not comparable to AudioLDM2/AudioGen as those are trained with AudioSet. \textbf{$^\dagger$} Stable Audio models are trained to generate longer outputs, but during inference can generate variable-length outputs relying on the timing conditioning. We trim the generated audio to 10 sec (discarding the end silent part) for a fair comparison.}
\label{tab:audiocaps}
\end{table*}





\begin{table*}[]
\centering
\begin{tabular}{lccccc}
\toprule
                  &             & output &                          &                         &                                \\ 
                  & channels/sr & length & FD$_{openl3} \downarrow$ & KL$_{passt} \downarrow$ & CLAP$_{score } \uparrow$           \\ \midrule
MusicGen-large-stereo~\cite{copet2023simple} &   2/32kHz          &   47   &     190.47   &     0.52    &   0.31     \\
{Stable Audio 1.0}~\cite{stableaudio}   & {2}/44.1kHz & 95 sec $^\dagger$ & 142.50  &   0.40 & 0.38 \\
{Stable Audio 2.0}~\cite{evans2024long}    & {2}/44.1kHz & 190 sec $^\dagger$ & \textbf{71.25} & \textbf{0.37} & \textbf{0.42}  \\
{Stable Audio 2.0}~\cite{evans2024long}   & {2}/44.1kHz & 285 sec $^\dagger$ & 81.05 & 0.39 & 0.42  \\
Stable Audio Open & {2}/44.1kHz &  47  & 96.51 & 0.55 & 0.41   \\ \bottomrule
\end{tabular}
\caption{\textit{Song Desciber Dataset (instrumental music, not sounds)}. Stable Audio Open is worse than Stable Audio at generating music, but slightly better than MusicGen (best open model). \textbf{$^\dagger$} Stable Audio models are trained to generate longer outputs, but during inference can generate variable-length outputs relying on the timing conditioning. We trim the generated audio to 47 sec (discarding the end silent part) for a fair comparison.}
\label{tab:sdd_nosinging}
\end{table*}

\begin{table*}[]
\centering
\begin{tabular}{lcccccc}
\toprule
                                                          & sampling & STFT & MEL &  & latent    &   latent  \\   
                                                         &rate & distance $\downarrow$ & distance $\downarrow$ & SI-SDR $\uparrow$  &  rate&     (channels)\\ \midrule                                                            

DAC~\cite{DAC} & 44.1kHz &  1.04 & 0.63   &  5.05  &     86Hz &     discrete\\ 
AudioGen~\cite{audiogen} & 48kHz   &  1.10  &  0.62 &  4.60   &     50Hz &     discrete\\  
Encodec~\cite{Encodec,copet2023simple} & 32kHz  & 1.82   & 1.12 &5.33  &    50Hz &     discrete\\ \midrule \midrule 
AudioGen~\cite{audiogen} & 48kHz &   1.22  & 0.73  &  2.04    &     100Hz &     continuous (32)\\ 
{Stable Audio} 1.0~\cite{stableaudio} & 44.1kHz & 1.32   &  0.86  &   1.81    &     43Hz &     continuous (64)\\  \midrule 
Stable Audio 2.0~\cite{evans2024long} & 44.1kHz  &  \textbf{1.21} &  \textbf{0.85}  &   \textbf{0.33}  &     21.5Hz&     continuous (64) \\ 
Ours & 44.1kHz  &  1.25  &  0.86  &    -0.93  &     21.5Hz&     continuous (64) \\ 
\bottomrule
\end{tabular}
\caption{\textit{Autoencoder reconstructions: AudioCaps Dataset (sounds)}. Different autoencoders use various sampling rates but evaluations are conducted at 44.1kHz for a fair comparison. Also note that (i) continuous latents are not comparable to discrete ones, and (ii) different latent rates are not strictly comparable. Sorted by latent rate.}
\label{tab:autoencoder_audiocaps}
\end{table*}










\begin{table*}[]
\centering
\begin{tabular}{lcccccc}
\toprule
                                                          & sampling & STFT & MEL &  & latent    &   latent  \\   
                                                         &rate & distance $\downarrow$ & distance $\downarrow$ & SI-SDR $\uparrow$  &  rate&     (channels)\\ \midrule                                                            

DAC~\cite{DAC} & 44.1kHz &   0.96  &  0.53  &  11.30     &     86Hz &     discrete\\ 
AudioGen~\cite{audiogen} & 48kHz   &   1.16 &  0.66 &  9.64  &     50Hz &     discrete\\  
Encodec~\cite{Encodec,copet2023simple} & 32kHz  &  1.70  & 1.09   &  5.75   &    50Hz &     discrete\\ \midrule \midrule 
AudioGen~\cite{audiogen} & 48kHz &  1.10   & 0.65    &  9.21    &     100Hz &     continuous (32)\\  
{Stable Audio 1.0}~\cite{stableaudio} & 44.1kHz & 1.20   &  0.67 & 9.12    &     43Hz &     continuous (64)\\  \midrule 
Stable Audio 2.0~\cite{evans2024long}  & 44.1kHz  &  \textbf{1.21}  & \textbf{0.72}  &  \textbf{7.65}    &     21.5Hz&     continuous (64) \\ 
Ours & 44.1kHz  &  1.29  &  0.77   & 6.28     &     21.5Hz&     continuous (64) \\ 
\bottomrule
\end{tabular}
\caption{\textit{Autoencoder reconstructions: Song Describer Dataset (instrumental music)}. Different autoencoders use various sampling rates but evaluations are conducted at 44.1kHz for a fair comparison. Also note that (i) continuous latents are not comparable to discrete ones, and (ii)~different latent rates are not strictly comparable. Sorted by latent rate.}
\label{tab:autoencoder_sdd_nosinging}
\end{table*}

\subsection{Prompts preparation for training the DiT}

Training audio is paired with text metadata.
Freesound examples include natural language descriptions as well as the title of the recording and tags. FMA music examples include metadata like year, genres, album, title, and artist.
We generate text prompts from the metadata by concatenating a random subset of the metadata as a string. This allows for specific properties to be specified during inference, while not requiring these properties to be present at all times. For metadata-types with a list of values, like tags or genres, we shuffle the list. As a result, we perform a variety of random transformations to the resulting string, shuffling the order and also transforming between upper and lower case. 
For half of the FMA prompts, we include the metadata-type (e.g., artist or album) and join them with a comma (e.g., \emph{``year: 2021, artist: dadabots, album: can't play instruments, title: pizza hangover"}).
For the other half, we do not include the metadata-type and join them with a comma (e.g., \emph{``dadabots, can't play instruments, pizza hangover, 2021"}). 

\section{Experimental Setup}\label{sec:training}

All models are trained with AdamW, with weight decay of 0.001 and a learning rate scheduler including exponential ramp-up and decay. The exact training hyperparameters we used are detailed online.\footnotemark[2]

\subsection{Autoencoder training}

The {(variational)} autoencoder is trained using a variety of objectives. First, with a reconstruction loss, based on a perceptually weighted multi-resolution STFT~\cite{auraloss} dealing with stereo audio via the mid-side (M/S) and left-right channels (L/R) representations. The L/R component is down-weighted by 0.5 compared to the M/S component, and exists to mitigate ambiguity around left-right placement. Second, we employ an adversarial loss term with feature matching, utilizing 5 convolutional discriminators as in Encodec~\cite{Encodec}. 
Third, we use a KL divergence loss term that is down-weighted by 1$\times 10^{-4}$.

It trained for 183 h on $\times$32 A100s with a batch size of 4. At this point the encoder was frozen, and the decoder was trained for another 273 h on $\times$32 A100s with a batch size of 8. Each batch is made of $\approx$1.5 sec chunks (65,536 samples at 44.1kHz). The autoencoder itself was trained with a base learning rate of 1.5$\times 10^{-4}$, and the discriminators with a learning rate of 3$\times 10^{-4}$.

\subsection{DiT training and inference}

The DiT is trained to predict a noise increment from noised ground-truth latents, following the v-objective~\cite{salimans2022progressive} approach. During inference, we use the DPM-Solver++~\cite{lu2022dpm} {for 100 steps} with classifier-free guidance (scale of 7.0). The DiT is trained for 338~h on $\times$64 A100s with a batch size of 4 and a base learning rate of 5$\times 10^{-5}$. Each batch contains latent sequences of length 1024 ($\approx$47 sec).


\section{Evaluation}
\label{sec:evaluation}

\subsection{Generative model evaluation} 

We employ established quality metrics\footnote[3]{https://github.com/Stability-AI/stable-audio-metrics} that include FD$_{openl3}$~\cite{cramer2019look}, KL$_{passt}$~\cite{koutini22passt} and CLAP$_{score}$~\cite{wu2023large,huang2023make}. A low FD$_{openl3}$ implies that the generated audio is plausible and closely matches the reference~\cite{kilgour2018fr,copet2023simple}. A low KL$_{passt}$ indicates semantic correspondence between the generated and the reference audio~\cite{copet2023simple}. A high CLAP$_{score}$ denotes that the generated audio adheres to the given text prompt~\cite{wu2023large,huang2023make}. 
We use two evaluation sets: AudioCaps Dataset~\cite{kim2019audiocaps} for sound generation, and Song Describer Dataset~\cite{manco2023song} for music generation.

Table~\ref{tab:audiocaps} shows the AudioCaps Dataset results. AudioCaps' test set contains 979 audio segments from YouTube, each with several captions (881 audios were available and it includes 4,875 captions).
We generate an audio per caption, resulting in 4,875 generations. 
We use previous Stable Audio~\cite{stableaudio, evans2024long} models as baselines. 
We also include AudioLDM2~\cite{liu2023audioldm2} and AudioGen~\cite{audiogen} as a reference, since they are the most competitive open models available, but those might not be strictly comparable since they were trained on AudioSet (and AudioCaps is a subset of AudioSet).
Stable Audio Open outperforms comparable baselines, particularly on FD$_{openl3}$, highlighting its potential to generate realistic sounds and field recordings.

Table~\ref{tab:sdd_nosinging} shows the Song Describer Dataset results. We select this benchmark because others contain bad quality music and shorter snippets~\cite{agostinelli2023musiclm}. As vocal generation is not our focus and baselines are not trained for that, we run a fair evaluation using a subset of prompts describing instrumental music (no-singing) \cite{stableaudio,evans2024long} with 586 captions for 446 music tracks.  We generate an audio per caption, which results in 586 generations. 
We set previous Stable Audio~\cite{stableaudio, evans2024long} models and MusicGen-large-stereo~\cite{copet2023simple} as baselines. The latter is the most competitive open model for stereo music generation~\cite{stableaudio}. Our results suggest that Stable Audio Open is worse than Stable Audio at generating music but slightly better than MusicGen (best open model).

\subsection{Autoencoder evaluation}

Note that the autoencoder can also be used alone and is implicitly released with the model. For this reason we also evaluate its audio reconstruction quality (Tables \ref{tab:autoencoder_audiocaps} and \ref{tab:autoencoder_sdd_nosinging}) by comparing ground-truth and reconstructed audio via a set of established audio quality metrics~\cite{DAC,Encodec}: STFT distance, MEL distance and SI-SDR (as in auraloss library~\cite{auraloss}, with its default parameters). The reconstructed audio is obtained by encoding-decoding the ground-truth audio from the AudioCaps Dataset (881 recordings) and Song Describer Dataset (no-singing subset with 446 tracks). We compare our autoencoder against a number of neural audio codecs including Encodec~\cite{Encodec}, DAC~\cite{DAC} and AudioGen~\cite{audiogen}.
We select the Encodec 32kHz variant because the MusicGen-large-stereo baseline relies on it, and DAC 44.1kHz because its alternatives operate at 24kHz and 16kHz. 
Further, as our autoencoder relies on a continuous latent, we also compare with AudioGen, a state-of-the-art autoencoder with both continuous and discrete latent options.
All our baselines are stereo, except DAC and Encodec. In those cases, we independently project left and right channels and reconstruct from those. 
Results show that lower latent rates generally yield worse reconstructions.
Considering this, note that our model is only comparable to Stable Audio 2.0 (being continuous, with the same latent rate) and {shows nearly similar performance despite being trained only with CC data.}

\subsection{Memorization (exact copies) analysis}

Recent works~\cite{evans2024long, carlini2023extracting, esser2024scaling} examined the potential of generative models to memorize training data, especially for repeated elements in the training set. Further, MusicLM~\cite{agostinelli2023musiclm} conducted a memorization analysis to address concerns on the potential misappropriation of creative content. Adhering to principles of responsible model development, we also run a comprehensive study on memorization~\cite{agostinelli2023musiclm,evans2024long, carlini2023extracting,esser2024scaling}. 

In light of the possible risk of memorizing repeated audio within the training set, we start by studying if our dataset contains repeated data.
We embed all our training data using the LAION-CLAP~\cite{wu2023large} audio encoder to select audios that are close in this space based on a manually set threshold. The threshold is set such that the selected audio correspond to exact replicas.
With this process, we identify 3,693 Freesound and 856 FMA repeated recordings.

Our methodology is based on comparing our model's generations against the training set in LAION-CLAP space. We then select the top-50 generations that are closest to the training data (the memorization candidates) and listen. 
We listened to memorization candidates generated with prompts from the identified repeated data in our training set, and did not find memorization.
We also listened to memorization candidates from 11,000 random prompts from the training set, and did not find memorization.
We even listened to memorization candidates from outstanding generations, and did not find memorization. 
The most interesting memorization candidates, together with their closest training data, are online for listening\footnote[4]{https://stability-ai.github.io/stable-audio-open-demo/}. Those include similar generations of well defined sounds like \textit{``storm"} or \textit{``1000Hz"}, but we did not find memorization beyond that.

Also note that our model only generates 47s audio and cannot memorize longer audio. Further, it cannot produce intelligible speech or singing, making it difficult to memorize such examples.

Finally, note that the primary objective of generative modeling is to create new content based on the training data. Simply reproducing the training data indicates poor performance and is not interesting.

\subsection{Inference speed on various hardware}

Our model runs 8 inference steps per second (steps/sec) on an RTX-3090 (24GB VRAM), 11 steps/sec on an RTX-A6000 (48GB), and 20 steps/sec on an H100 (80GB). Appendices \ref{app:memory} and \ref{app:finetuning} include more details on VRAM usage and on which hardware it can be finetuned.

\section{Conclusions}

This article documents the release of Stable Audio Open with a particular focus on evaluation and data transparency. Our results show its potential for synthesizing high-quality stereo sounds at 44.1kHz. Further, we also release a continuous autoencoder that operates at a low latent rate (21.5Hz) that can work for both music and audio. Our model is accessible to everyone\footnotemark[2] and can run on consumer-grade GPUs, making it appealing for both academic and artistic use cases. 

\subsection{Limitations}
\label{sec:limitations}

\hspace{5mm}{{\textit{On audio generation}}}. Our model finds challenging to generate prompts with {connectors}\footnote[5]{Including: \textit{and}, \textit{followed}, \textit{while}, \textit{as}, \textit{then}, \textit{with}, \textit{later}, \textit{before}, \textit{after}.} and cannot generate intelligible {speech}\footnote[6]{Including: \textit{speech}, \textit{male}, \textit{female}, \textit{woman}, \textit{man}, \textit{speaking}, \textit{speaks}.}. It sometimes omits one (or more) of the sounds present in prompts with connectors, e.g., \emph{``A man speaking as a crowd of people laugh and applaud"} where the generated sound includes a man speaking with mild background noise, but with no laughter or applause.
Also, speech generations are not intelligible because our model is not ``spoken-word" conditioned. 
Table \ref{tab:claps} shows that $\text{CLAP}_{score}$ improves when removing {connectors}- and {speech}-related prompts. 

\begin{table}[h]
\centering
\begin{tabular}{l|cc}
\toprule
             & $\text{CLAP}_{score}$ ($\uparrow$) & \# files\\ \midrule
AudioCaps    &     0.29 & 4,875 \\
-- \textit{no speech prompts} & 0.31 & 2,765\\
-- \textit{no connectors prompts} &  0.32 & 711 \\
-- \textit{no speech \& connectors}    &  0.34 & 587 \\
\bottomrule
\end{tabular}
\caption{$\text{CLAP}_{score}$ depending on the used prompts.}
\label{tab:claps}
\end{table}

{\textit{On music generation}}.
Note that most commercial music is copyrighted. Hence, our model was trained with limited high-quality music since we focused on CC training data.
As a result, it is not competitive against state-of-the-art music models (Tables \ref{tab:sdd_nosinging} and \ref{tab:autoencoder_sdd_nosinging}).

{\textit{On prompting}}. Due to the above limitations, prompt engineering may be required for best results. Further, it was mainly trained with English text and is not expected to perform well in other languages.

\newpage

\bibliographystyle{IEEEbib}
\bibliography{strings}

\appendix

\begin{table*}[h]
\vspace{-4mm}
\centering
\begin{tabular}{lccccc}
\toprule
                  &             & output &                          &                         &                                \\ 
                  & channels/sr & length & FD$_{openl3} \downarrow$ & KL$_{passt} \downarrow$ & CLAP$_{score } \uparrow$           \\ \midrule
MusicGen-large-stereo~\cite{copet2023simple} &   2/32kHz          &   47   &               193.98     &             0.54        &           0.30                   \\
{Stable Audio 1.0}~\cite{stableaudio}   & {2}/44.1kHz & 95 sec $^\dagger$ & 139.41 & 0.36 & 0.40  \\
{Stable Audio 2.0}~\cite{evans2024long}    & {2}/44.1kHz & 190 sec $^\dagger$ &  \textbf{72.17} & \textbf{0.35} & \textbf{0.44}  \\
{Stable Audio 2.0}~\cite{evans2024long}   & {2}/44.1kHz & 285 sec $^\dagger$ &  80.44& 0.36 & 0.43  \\
Stable Audio Open & {2}/44.1kHz &  47  & 99.70 & 0.62 & 0.43             \\ \bottomrule
\end{tabular}
\caption{\textit{Song Desciber Dataset ({all dataset} with music, not sounds)}. Stable Audio Open is worse than Stable Audio at generating music, but slightly better than MusicGen (best open model). \textbf{$^\dagger$} Stable Audio models are trained to generate longer outputs, but during inference can generate variable-length outputs relying on the timing conditioning. We trim the generated audio to 47 sec (discarding the end silent part) for a fair comparison.}
\label{tab:sdd_appendix}
\end{table*}

\begin{table*}[h!]
\centering
\vspace{-0mm}
\begin{tabular}{lcccccc}
\toprule
                                                          & sampling & STFT & MEL &  & latent    &   latent  \\   
                                                         &rate & distance $\downarrow$ & distance $\downarrow$ & SI-SDR $\uparrow$  &  rate&     (channels)\\ \midrule                                                            

DAC~\cite{DAC} & 44.1kHz & 0.96    & 0.52   &     10.83  &     86Hz &     discrete\\ 
AudioGen~\cite{audiogen} & 48kHz   &  1.17   & 0.64   &     9.27  &     50Hz &     discrete\\  
Encodec~\cite{Encodec,copet2023simple} & 32kHz  &  1.82   & 1.12   &    5.33  &    50Hz &     discrete\\ \midrule \midrule 
AudioGen~\cite{audiogen} & 48kHz &  1.10    & 0.64   &     8.82  &     100Hz &     continuous (32)\\  
{Stable Audio 1.0}~\cite{stableaudio} & 44.1kHz &  1.19    & 0.67    &     8.62   &     43Hz &     continuous (64)\\ \midrule
Stable Audio 2.0~\cite{evans2024long}  & 44.1kHz  &  \textbf{1.19}    & \textbf{0.71}    &     \textbf{7.14}  &     21.5Hz&     continuous (64) \\ 
Ours & 44.1kHz  &  1.32  & 0.78    &  5.78    &     21.5Hz&     continuous (64) \\ 
\bottomrule
\end{tabular}
\caption{\textit{Autoencoder reconstructions: Song Describer Dataset ({all dataset})}. Different autoencoders use various sampling rates but evaluations are conducted at 44.1kHz for a fair comparison. Also note that (i) continuous latents are not comparable to discrete ones, and (ii)~different latent rates are not strictly comparable. Sorted by latent rate.}
\label{tab:autoencoder_sdd_appendix}
\end{table*}

\newpage

\section{Additional Song Describer Dataset results}
\label{sec:appendix}

Tables~\ref{tab:sdd_nosinging} and \ref{tab:autoencoder_sdd_nosinging} show the results for a subset of the Song Describer Dataset that includes only prompts for instrumental music\footnote{Prompts containing any of those words were removed: \textit{speech, speech synthesizer, hubbub, babble, singing, male, man, female, woman, child, kid, synthetic singing, choir, chant, mantra, rapping, humming, groan, grunt, vocal, vocalist, singer, voice, and acapella}.}~\cite{stableaudio,evans2024long}. As vocal generation is not our focus and baselines were not trained for that either, the main body of the paper reports results on the instrumental subset to ensure a fair evaluation. Yet, for completeness, this appendix also includes results on the entire Song Describer Dataset (Tables~\ref{tab:sdd_appendix} and~\ref{tab:autoencoder_sdd_appendix}) with the same metrics as in Tables~\ref{tab:sdd_nosinging} and \ref{tab:autoencoder_sdd_nosinging}. Note that although results vary slightly, the trends remain consistent.

\subsection{Generative model evaluation}
Table~\ref{tab:sdd_appendix} shows the Song Describer Dataset (all dataset) results. The dataset contains 1,106 captions for 706 tracks~\cite{manco2023song}.  We generate an audio per caption, which results in 1,106 generations. Our results suggest that Stable Audio Open is worse than Stable Audio at generating music, but slightly better than MusicGen (best open model).

\subsection{Autoencoder evaluation}

Table~\ref{tab:autoencoder_sdd_appendix} shows the Song Describer Dataset (all dataset) results. 
We evaluate the reconstructed audio obtained by encoding-decoding the 706 tracks~\cite{manco2023song} from
the dataset.
Results show that lower latent rates generally yield worse reconstructions.
Considering this, note that our model is only comparable to Stable Audio 2.0 (being continuous, with the same latent rate) and shows slightly worse performance despite being trained only with CC data.

\newpage

\section{VRAM consumption during inference}
\label{app:memory}

 During diffusion the DiT utilizes 5.9 GB VRAM. During decoding, rendering waveforms from latents, RAM usage increases to 14.5 GB.


\subsection{Chunk decoding}

Decoding waveforms from latents consumes significantly more VRAM than the DiT. To reduce the memory footprint of the decoder, we explore chunk decoding by splitting the latent sequence into overlapping chunks, decoding them separately, and reassembling them into a final audio. As long as overlaps include the decoder's receptive field (16 latents on each side), the resulting audio is the same. The VRAM usage after chunking is as follows:

\begin{figure}[h]
    \centering
    \includegraphics[width=1\linewidth]{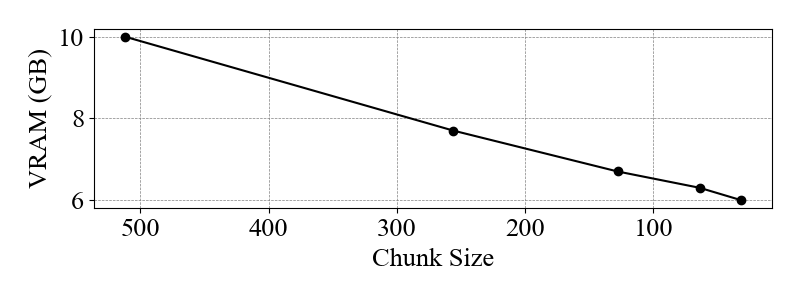}
\vspace{-11mm}    
\end{figure}

\section{Finetuning on various hardware}
\label{app:finetuning}

\textit{@RoyalCities} changed the window length from 47 to 20 sec, and finetuned it on piano loops using $\times$2 RTX-A6000 (48GB)\footnote{https://x.com/RoyalCities/status/1808563794677018694}. 
\textit{@\_lyraaaa\_} changed the window length from 47 to 11 sec, and fine-tuned it on loops/oneshots using $\times$1 or $\times$2 RTX-A6000 (48GB)\footnote{https://www.youtube.com/watch?v=ex4OBD\_lrds}.

\end{document}